\documentclass[conference,10pt]{IEEEtran}
\usepackage{epsfig,rotating,setspace,latexsym,amsmath,epsf,amssymb,amsfonts,bm,theorem,subfigure,epstopdf}
\usepackage{cite,authblk}
\usepackage{bbm}
\usepackage{algorithmic,algorithm}
\newtheorem{theorem}{Theorem}
\newtheorem{lemma}{Lemma}
\newenvironment{Proof}[1]{\medskip\par\noindent{\bf Proof:\,}\,#1}{{\mbox{\,$\blacksquare$}\par}}
\IEEEoverridecommandlockouts

\allowdisplaybreaks
\usepackage{color}
\usepackage{graphicx}

\begin{document}
	
\title{Age of Information Scaling in Large Networks\thanks{This work was supported by NSF Grants CNS 15-26608, CCF 17-13977 and ECCS 18-07348.}}
	
\author[1]{Baturalp Buyukates}
\author[2]{Alkan Soysal}
\author[1]{Sennur Ulukus}
\affil[1]{\normalsize Department of Electrical and Computer Engineering, University of Maryland, MD}
\affil[2]{\normalsize Department of Electrical and Electronics Engineering, Bahcesehir University, Istanbul, Turkey}
	
\maketitle
	
\begin{abstract}
We study age of information in a multiple source-multiple destination setting with a focus on its scaling in large wireless networks. There are $n$ nodes that are randomly paired with each other on a fixed area to form $n$ source-destination (S-D) pairs. We propose a three-phase transmission scheme which utilizes local cooperation between the nodes by forming what we call \emph{mega update packets} to serve multiple S-D pairs at once. We show that under the proposed scheme average age of an S-D pair scales as $O(n^{\frac{1}{4}})$ as the number of users, $n$, in the network grows. To the best of our knowledge, this is the best age scaling result for a multiple source-multiple destination setting.
\end{abstract}

\section{Introduction}
Following the pioneering work of Gupta and Kumar \cite{Gupta00}, scaling laws for \emph{throughput} have been extensively studied in the literature, e.g., \cite{Gamal06a, Gamal06b, Grossglauser02, Neely05, Sharma06} which studied dense and extended networks considering static and mobile nodes. This literature that started with Gupta-Kumar's multi-hop scheme that achieved a total throughput of $O(\sqrt{n})$ for the network, and hence, $O(\frac{1}{\sqrt{n}})$ throughput per-user, has culminated in the seminal papers of Ozgur et al. \cite{Ayfer07, Ayfer10} which achieved a total throughput of $O(n)$ for the network, and hence, $O(1)$ throughput per-user, by making use of \emph{hierarchical cooperation} between nodes. In this paper, we study scaling of \emph{age of information} in large wireless networks.

Age of information is a recently proposed metric that measures the freshness of the received information. A typical model to study age of information includes a source which acquires time-stamped status updates from a physical phenomenon. These updates are transmitted over a network to the receiver(s) and age of information in this network or simply the $age$ is the time elapsed since the most recent update at the receiver was generated at the transmitter. In other words, at time $t$, age $\Delta(t)$ of a packet which was generated at time $u(t)$ is $\Delta(t) = t-u(t)$. Age of information as the freshness metric of a system has been extensively studied in a queueing-theoretic setting in references \cite{Kaul12b, Costa14, Huang15, Yates15, Bedewy16, Sun16b, Najm17, Kadota16} and in an energy harvesting setting in references \cite{Arafa17b, Arafa17a, Bacinoglu15, Wu18, Arafa18a, Arafa18b, Arafa18d, Baknina18a, Baknina18b}.

Considering dense IoT deployments and the increase in the number of users in networks supplying time-sensitive information, the scalability of age as a function of the number of nodes has become a critical issue. What makes age analysis in a large network challenging is the fact that good age performance corresponds to neither high throughput nor low delay. As argued in references \cite{Kadota16} and \cite{Pappas15}, for optimized timeliness of the updates, we need regular packet delivery with low delay. Maximum throughput can be achieved by sending as many updates as possible from the source. However, this may cause congestion in the system resulting in stale packet deliveries at the destination. Likewise, delay in the network can be reduced by decreasing the update rate which results in outdated information at the destination since the update delivery frequency is low. In this paper, we balance these two opposing objectives, and develop an achievable scheme that strikes a balance between the two in large networks.

References that are most closely related to our work are \cite{Ioannidis09, Zhong17a, Buyukates18, Jiang18a}. Reference \cite{Ioannidis09} studies a mobile social network with a single service provider and $n$ communicating users, and shows that under Poisson \emph{contact processes} among users and uniform rate allocation from the service provider, the average age of the content at the users is $O(\log n)$. They also show that without the contact process between the users, age grows linearly in $n$, as the service provider serves only one user at a time. Reference \cite{Zhong17a} studies the age scaling in a multicast setting and shows that if the source waits for the earliest $k$ of the total $n$ nodes to successfully receive the update during the multicast before sending the next one, a constant age scaling can be achieved. In \cite{Buyukates18}, we extend this result to two-hop multicast networks showing that if earliest $k_1$ and earliest $k_2$ schemes are adapted in the first and second hops, respectively, we can again achieve a constant average age scaling at the end nodes. Note that all these works focus on a setting in which there is a single source that updates multiple destination nodes.

In this work, we focus on a multiple source-multiple destination setting and study a network of $n$ nodes on a fixed area that want to communicate with each other. Each node serves both as a source and a destination. In other words, nodes are paired randomly and we have $n$ source-destination (S-D) pairs to serve. Our goal is to find a tranmission scheme which allows all $n$ S-D pairs to communicate and achieve the smallest average age scaling at the destination nodes.

As studied in \cite{Jiang18a}, a straightforward way to achieve successful communication between all S-D pairs is to use a round-robin policy such that at each turn only one source transmits to its destination and stays silent while all other sources transmit successively during their respective turns. This direct method achieves an age scaling of $O(n)$ meaning that age increases linearly in $n$ since under this policy average inter-update times at a destination node increases linearly as $n$ grows making the updates less frequent and causing age to increase.

As in the setting of \cite{Gupta00}, a multihop scheme that involves successive transmissions between the source and destination nodes can be utilized. In that work the network is divided into cells such that each cell contains at least one node. Their scheme involves hops between these cells so that $O(\sqrt{n})$ messages are carried by each cell. Each of these cells can be considered a queue with multiple sources. As studied in \cite{Yates12}, age of a single update packet that is served by a queue with $O(\sqrt{n})$ different packet streams is also $O(\sqrt{n})$ under LCFS with preemption policy. This means that using multihop scheme after one hop age of an update packet becomes $O(\sqrt{n})$ since the queue is shared by many other packets. Considering the fact that the number of hops needed is $O(\sqrt{n})$, using a multihop scheme, the average age scales as $O(n)$ as in \cite{Jiang18a}.

As studied in reference \cite{Ayfer07}, a more complicated scheme involving hierarchical cooperation between the users locally and MIMO transmissions across the network can be employed. Although this approach gives the optimal throughput scaling in a large network, in \cite{Ayfer10} authors also show that their original hierarchical scheme has a poor delay performance, in that, it scales as $O(n^{\frac{h}{h+1}})$, where $h$ is the number of hierarchy levels. When we have just one hierarchy level, i.e., $h=1$, the delay scaling is $O(\sqrt{n})$, which is the same as that of a multihop scheme, as shown in \cite{Gamal06a}. However, as we increase $h$ to achieve better throughput, delay performance tends to $O(n)$.

In this paper, considering all these previous results, we propose a three-phase transmission scheme to serve all $n$ S-D pairs such that time average age of each node is small. Our scheme utilizes local cooperation between the users as in \cite{Ayfer07}. We divide the network into cells of equal number of users. In the first phase, nodes from the same cell communicate to create a \emph{mega update packet}, which contains the updates of all nodes from that cell. In the second phase, inter-cell communication takes place and each cell sends its mega packet to the corresponding destination cells. The main idea behind the mega update packets is to serve many nodes at once to decrease inter update time. In the third and final phase, received mega update packet is relayed to the intended recipient node in the cell. During all these phases, we make use of the spatial separation of the nodes to allow multiple simultaneous transmissions provided that there is no destructive interference caused by others. Using this scheme, we achieve an age scaling of $O(n^{\frac{1}{4}})$ per-user, which to the best of our knowledge, is the best age scaling result in a multiple source-multiple destination setting. We note that we do not utilize any hierarchy as \cite{Ayfer10} shows that it gives a poor delay performance.

\section{System Model and Age Metric} \label{model}

We consider $n$ nodes that are uniformly and independently distributed on a square of fixed area $S$. Every node is both a source and a destination. These sources and destinations are paired randomly irrespective of their locations. Thus, for a total of $n$ nodes in the network, we have $n$ S-D pairs. Sources create update packets and transmit them to their respective destinations using the common wireless channel. Each source has the same traffic rate and transmit power level. Each source wants to keep its destination as up-to-date as possible. This makes each S-D pair need regular updates sent with low transmission delay and hence brings up the concept of age of information. Age is measured for each destination node and for node $i$ at time $t$ age is the random process $\Delta_i(t) = t - u_i(t)$ where $u_i(t)$ is the timestamp of the most recent update at that node. The metric we use, time averaged age, is given by,
\begin{align}
\Delta_i = \lim_{\tau\to\infty} \frac{1}{\tau} \int_{0}^{\tau} \Delta_i(t) dt \label{avg_age}
\end{align}
for node $i$. We will use a graphical argument similar to \cite{Buyukates18} to derive the average age for a single S-D pair.

Inspired by \cite{Ayfer07}, we will propose a scheme based on clustering nodes and making use of what we call \emph{mega update packets} to increase the spatial reuse of the common wireless channel. This entails dividing $n$ users into $\frac{n}{M}$ cells with $M$ users in each cell with high probability. The users communicate locally within cells to form the mega update packets. We model the delay in these intra-cell communications as i.i.d.~exponential with parameter $\lambda$. Then, mega packets are transmitted between the cells. We model the delay in these inter-cell communications as i.i.d.~exponential with parameter $\tilde{\lambda}$. Finally, the individual updates are extracted from mega updates and distributed to the intended destinations within cells again via intra-cell communications. While intra-cell communications occur simultaneously in parallel across the cells (see Section~\ref{note_phaseI} for details), inter-cell updates occur sequentially one-at-a-time.

Due to i.i.d.~nature of service times in intra- and inter-cell communications, all destination nodes experience statistically identical age processes and will have the same average age. Thus, we will drop user index $i$ in (\ref{avg_age}) and use $\Delta$ instead of $\Delta_i$ in the ensuing analysis.

Finally, we denote the $k$th order statistic of random variables $X_1, \ldots ,X_n$ as $X_{k:n}$. Here, $X_{k:n}$ is the $k$th smallest random variable, e.g., $X_{1:n}=\min\{X_i\}$ and $X_{n:n}=\max\{X_i\}$. For i.i.d.~exponential random variables $X_i$ with parameter $\lambda$,
\begin{align}
E[X_{k:n}] =& \frac{1}{\lambda}(H_n - H_{n-k}) \\
Var[X_{k:n}] =& \frac{1}{\lambda^2}(G_{n} - G_{n-k})
\end{align}
where $H_n = \sum_{j=1}^{n} \frac{1}{j}$ and $G_n = \sum_{j=1}^{n} \frac{1}{j^2}$. Using these,
\begin{align}
E[X_{k:n}^2] =&  \frac{1}{\lambda^2}\left((H_n - H_{n-k})^2 + G_{n} - G_{n-k} \right)
\end{align}

\section{Age Analysis of a Single S-D Pair} \label{subsection:age}

The network operates in sessions such that during each session all $n$ sources successfully send their update packets to their corresponding destinations. Each session lasts for $Y$ units of time. Here, we calculate the age of a single S-D pair $(s_i,d_i)$ considering a generic \emph{session time} $Y$. In the next section when the proposed scheme is presented, a specific session time will be characterized. As explained in Section~\ref{model}, it is sufficient to analyze the age of a single S-D pair since each pair experiences statistically identical ages.

\begin{figure}[t]
	\centering  \includegraphics[width=0.885\columnwidth]{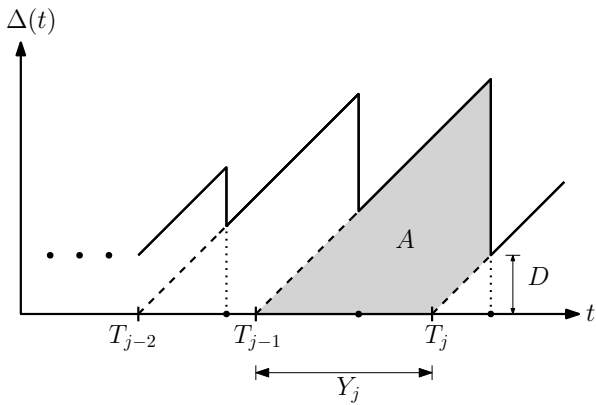}
	\caption{Sample age $\Delta(t)$ evolution for a single S-D pair. Update deliveries  are shown with $\bullet$. Session $j$ starts at time $T_{j-1}$ and lasts until $T_j = Y_j + T_{j-1}$.}
	\label{fig:ageEvol}
	\vspace{-0.5cm}
\end{figure}

Session $j$ starts at time $T_{j-1}$ and all sources including $s_i$ generate their respective $j$th update packets. This session lasts until time $T_j = T_{j-1}+Y_j$, at which, all $n$ destinations successfully receive their updates from their corresponding sources. In other words, a session ends when the last S-D pair finishes the packet transmission. Thus, a destination can receive its update packet before the session ends. Fig.~\ref{fig:ageEvol} shows the evolution of the age at a destination node over time. It is in the usual sawtooth shape with the age increasing linearly over time and dropping to a smaller value as the updates are received at the destination. The process repeats itself at time $T_j$ when all sources including $s_i$ generate the next update packet, namely update $j+1$.

Using Fig.~\ref{fig:ageEvol} the average age for an S-D pair is given by
\begin{align}
\Delta =& \frac{E[A]}{E[L]}  \label{age_formula}
\end{align}
where $A$ denotes the shaded area and $L$ is its length. From the figure, we observe that $E[L] = E[Y]$ and $E[A] = \frac{E[Y^2]}{2}+E[D]E[Y]$. Using these in (\ref{age_formula}) we obtain,
\begin{align}
\Delta =& E[D]+ \frac{E[Y^2]}{2E[Y]} \label{age}
\end{align}
where $D$ denotes the time from the generation of an update till its arrival at the destination. Note that in some systems $D$ may be directly equal to the link delay. However, as in our model here, $D$ may capture some additional delays that may occur during the service time of an update. This will be further clarified in the next section.

\section{Proposed Transmission Scheme} \label{section:scheme}

Proposed scheme involves clustering nodes and making use of \emph{mega update packets} to serve many S-D pairs at once to reduce the session time. In this section, we describe the proposed three-phase transmission scheme and define mega update packets. As in \cite{Ayfer07}, we divide the square network area into $\frac{n}{M}$ cells of equal area such that each cell includes $M$ nodes with high probability which tends to $1$ as $n$ increases. The transmission delays between the nodes belonging to the same cell are denoted by $X_i$ whereas the transmission delays between the nodes from different cells are denoted by $\tilde{X}_i$. Note that $X_i$ and $\tilde{X}_i$ are independent; $X_i$ are i.i.d.~exponential with parameter $\lambda$ and $\tilde{X}_i$ are i.i.d.~exponential with parameter $\tilde{\lambda}$.

\begin{figure}[t]
	\centering  \includegraphics[width=0.8\columnwidth]{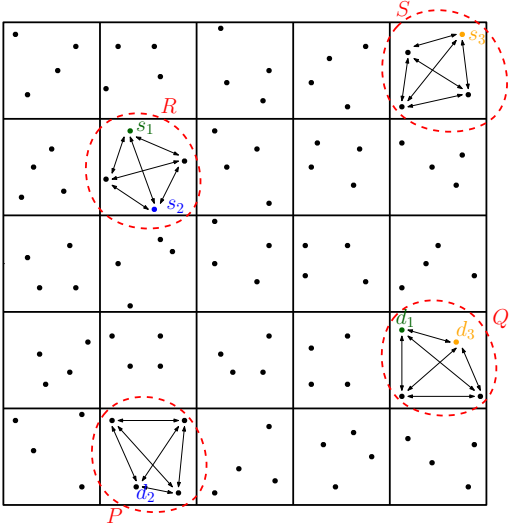}
	\caption{Cell formation for $M=4$ and $n=100$. Simultaneous intra-cell transmissions are depicted for three S-D pairs from cells $P$, $Q$, $R$ and $S$.}
	\label{fig:Phase1}
	\vspace{-0.5cm}
\end{figure}

{\bf Phase~I. Creating mega update packets:} In a cell, each one of the $M$ nodes distributes its current update packet to remaining $M-1$ nodes. This operation resembles the wait-for-all scheme studied in \cite{Zhong17a} since each node keeps transmitting until all $M-1$ nodes receive its packet. Thus, the time needed for each node to distribute its update packet to other nodes in the cell is $U = X_{M-1:M-1}$. Considering $M$ successive transmissions for each node in the cell, this phase is completed in $V = \sum_{i=1}^{M} U_i$ units of time. By the end of this phase in a cell, each one of the $M$ nodes has $M$ different update packets one from each other node in that cell. Each node combines all these $M$ packets to create what we call a \emph{mega update packet} (see Fig.~\ref{fig:Phase1}). In order to reduce the session time, cells work in parallel during Phase~I (see Section~\ref{note_phaseI} for a detailed description of this operation). This phase ends when the slowest cell among simultaneously operating cells finishes creating its mega update packet. Phase~I takes $Y_I = V_{\frac{n}{M}:\frac{n}{M}}$ units of time, where $Y_I$ denotes the duration of Phase~I.

\begin{figure}[t]
	\centering  \includegraphics[width=1\columnwidth]{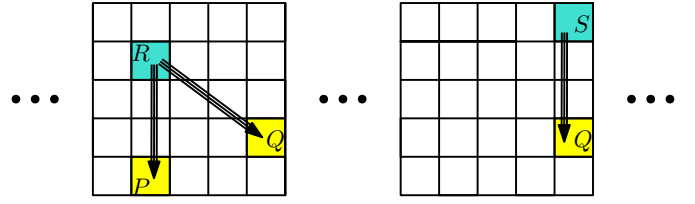}
	\caption{In Phase~II cells take turns to perform inter-cell transmissions. These inter-cell transmissions are shown for the same three S-D pairs depicted in Fig.~\ref{fig:Phase1}.}
	\label{fig:Phase2}
	\vspace{-0.5cm}
\end{figure}

{\bf Phase~II. MIMO-like transmissions:} In this phase, each cell successively performs MIMO-like transmissions using the mega update packets created in Phase~I. In each cell, all $M$ source nodes send the mega update packet through the channel at the same time to the respective destination cells in which the destination nodes are located. Since every node sends the same mega packet which includes all $M$ packets to be transmitted from that cell, this does not create interference. Thus, this is equivalent to sending update packets of all $M$ sources with $M$ copies each all at once (see Fig.~\ref{fig:Phase2}). Hence, this significantly reduces the time needed to transmit updates of all $M$ sources from that cell to their respective destinations. Note that this stage does not require the destination nodes to be in the same cell. In fact, considering that we have $M$ nodes in a cell, each cell can at most have $M$ different destination cells. Since we are sending $M$ copies of each update to a destination cell in which there are $M$ receiver nodes, only the earliest successful transmission is important. Thus, it takes $\tilde{U} = \tilde{X}_{1:M^2}$ units of time for a source node $s$ from cell $j$ to send its update to the destination cell where the destination node $d$ lies in. This MIMO-like transmissions of cell $j$ continues until the slowest source from that cell transmits its update. Hence, for each cell, this phase lasts for $\tilde{V} = \tilde{U}_{M:M}$. We repeat this for each cell, making the session time of this phase $Y_{II} =\sum_{i=1}^{\frac{n}{M}} \tilde{V}_i$.

{\bf Phase~III. In-cell relaying to the destination nodes:} By the end of Phase~II, each cell receives a mega packet for each one of its nodes. These packets may be received directly by their intended destination nodes. However, considering the worst case where they are received by any other node, we need to relay them to their actual designated recipient nodes. Thus, in this phase, all $M$ mega update packets received during Phase~II are sent to their recipients one at a time. Since this phase has intra-cell transmissions, it is performed in parallel across cells. For a single node this takes $X$ units of time, consequently we need $\hat{V} = \sum_{i=1}^{M} X_i$ to finish this process in a cell. As in Phase~I, we need to wait for the slowest cell to finish this phase. Then, $Y_{III} = \hat{V}_{\frac{n}{M}:\frac{n}{M}}$. Once the destination node receives the mega update packet it extracts the actual update sent from its own source.

The total session time of the proposed scheme is,
\begin{align}
Y = Y_I + Y_{II} + Y_{III} = V_{\frac{n}{M}:\frac{n}{M}}+ \sum_{i=1}^{\frac{n}{M}} \tilde{V}_i  + \hat{V}_{\frac{n}{M}:\frac{n}{M}} \label{sessiontime}
\end{align}
where $V$, $\tilde{V}$, and $\hat{V}$ are defined above. Note that in our proposed scheme we have,
\begin{align}
	D = Y_I + Y_{II} + Z \label{D}
\end{align}
where, as noted earlier, $D$ denotes the time between generation of an update at certain source node till its arrival at the corresponding destination node. Assuming no S-D pair is in the same cell, in our scheme, arrivals to destination nodes occur in Phase~III. This assumption is not critical because an S-D pair being in the same cell leads to a smaller $D$ and consequently to a smaller age. Therefore, by assuming no S-D pair is in the same cell, we essentially consider the worst case. Thus, any successful packet delivery will be no earlier than the duration of the first two phases $Y_I + Y_{II}$. On top of that, Phase~III involves $M$ successive in-cell transmissions for each node of a particular cell. Hence, depending on the cell that the source node lies in, as well as the realization of the transmission delay $X$, the corresponding destination node may receive the packet some time after Phase~III starts. For example, if a packet is the $j$th to be transmitted in Phase~III, then delivery will be at $Y_I+ Y_{II} + \sum_{i=1}^{j} {X}_i$. Then, the random variable $Z$ is of the form $Z=\sum_{i=1}^{j} {X}_i$.

Substituting (\ref{sessiontime})-(\ref{D}) in (\ref{age}) we obtain,
\begin{align}
\Delta =& 	E[Y_I] + E[Y_{II}] + E [Z] + \frac{E[Y^2] }{2E[Y]} \label{age_propScheme}
\end{align}
which is the average age of an S-D pair under the proposed transmission scheme.

Before we perform the explicit age calculation using (\ref{age_propScheme}), we make some observations to simplify our analysis. First, we note that, when the transmission delays $\tilde{X}$ are i.i.d.~exponential with rate $\tilde{\lambda}$, then $\tilde{U} = \tilde{X}_{1:M^2}$ is also exponential with rate $M^2\tilde{\lambda}$ \cite{Yates07}. Second, we have the following upper bound for the duration of Phase~I.

\begin{lemma} \label{lemma1}
$Y_I$ satisfies the following inequality,
\begin{align}
Y_I \leq \bar{V} \label{age_ineq}
\end{align}
where $\bar{V} = \sum_{i=1}^{M}\bar{U}_i$ and $\bar{U} = X_{n:n}$.
\end{lemma}

\begin{Proof}
Recall that $Y_I = V_{\frac{n}{M}:\frac{n}{M}}$, where $V = \sum_{i=1}^{M} U_i$ and $U = X_{M-1:M-1}$. To show the inequality we make the following observation: In Phase~I, $\frac{n}{M}$ cells operate simultaneously. First nodes of each of these cells start transmitting their packets to all other $M-1$ nodes of their cell at the same time. Since intra-cell transmission delays are all i.i.d.~across cells and packets, what we essentially have in this case is simultaneous transmission to $\frac{n}{M}(M-1)\approx n$ nodes, and therefore all first nodes will be done in $X_{n:n}$ units of time.
	
We repeat this for the second nodes of each cell and so on to get $\bar{V} = \sum_{i=1}^{M}(X_{n:n})_i = \sum_{i=1}^{M}\bar{U}_i$. In this way of operation, a cell waits for all other cells to finish distributing the update packet of the first node and then continues with the second node and so on. In a way, for each of its nodes it waits for the slowest cell to finish. However, in our constructed scheme during Phase I, inside a cell, nodes distribute their packets to other nodes of that cell without considering other cells and phase ends when all cells finish this process for all their $M$ nodes. Thus, $\bar{V}$ is an upper bound on $Y_I$.
\end{Proof}

Although our proposed Phase~I lasts shorter than the scheme described in Lemma~\ref{lemma1}, for tractability and ease of calculation we worsen our scheme in terms of session time, and take the upper bound in Lemma~\ref{lemma1} as our Phase~I duration such that from now on $Y_I = \bar{V}$. Third, we have the following upper bound for the duration of Phase~III.

\begin{lemma} \label{lemma2}
$Y_{III}$ satisfies the following inequality,
\begin{align}
Y_{III} \leq \bar{\bar{V}} \label{age_ineq2}
\end{align}
where $\bar{\bar{V}} = \sum_{i=1}^{M}\bar{\bar{U}}_i$ and $\bar{\bar{U}} = X_{\frac{n}{M}:\frac{n}{M}}$.
\end{lemma}

We omit the proof of Lemma~\ref{lemma2} since it follows similar to the proof of Lemma~\ref{lemma1}. Due to the same tractability issues, we worsen Phase~III as well in terms of duration and take $Y_{III} = \bar{\bar{V}}$ from now on.

As a result of above lemmas, (\ref{sessiontime}) becomes
\begin{align}
Y = \bar{V}+ \sum_{i=1}^{\frac{n}{M}} \tilde{V}_i + \bar{\bar{V}}\label{sessiontime2}.
\end{align}

Now, we are ready to drive an age expression using above lemmas in (\ref{age_propScheme}). This is stated in the following theorem.

\begin{theorem} \label{thm1}
Under the constructed transmission scheme, the average age of an S-D pair is given by,
	\begin{align}
	\Delta =& \frac{M}{\lambda} H_n + \frac{n}{M^3 \tilde{\lambda}} H_M + \frac{M-1}{2\lambda}H_{\frac{n}{M}} +\frac{1}{\lambda} \nonumber\\ &+  \frac{\frac{M^2}{\lambda^2}H_n^2+\frac{M}{\lambda^2}G_{n}}{2\left(\frac{M}{\lambda}H_n+\frac{n}{M^3\tilde{\lambda}}H_M +\frac{M}{\lambda}H_{\frac{n}{M}}\right)} \nonumber \\ &+ \frac{\frac{n^2}{M^6\tilde{\lambda}^2}H_M^2+\frac{n}{M^5\tilde{\lambda}^2}G_{M}}{2\left(\frac{M}{\lambda}H_n+\frac{n}{M^3\tilde{\lambda}}H_M +  \frac{M}{\lambda} H_{\frac{n}{M}}\right)} \nonumber \\ &+ \frac{\frac{M^2}{\lambda^2}H_{\frac{n}{M}}^2+\frac{M}{\lambda^2}G_{\frac{n}{M}}}{2\left(\frac{M}{\lambda}H_n+\frac{n}{M^3\tilde{\lambda}}H_M +\frac{M}{\lambda}H_{\frac{n}{M}}\right)} \nonumber \\ &+
	\frac{\frac{n}{M^2\lambda \tilde{\lambda}}H_n H_M + \frac{M^2}{\lambda^2}H_n H_{\frac{n}{M}} + \frac{n}{M^2 \lambda \tilde{\lambda}}H_M H_{\frac{n}{M}} }{\left(\frac{M}{\lambda}H_n+\frac{n}{M^3\tilde{\lambda}}H_M +\frac{M}{\lambda}H_{\frac{n}{M}}\right)}  \label{thm1res}	
	\end{align}
\end{theorem}

\begin{Proof}
The proof follows upon substituting (\ref{sessiontime2}) back in (\ref{age_propScheme}) and taking expectations of order statistics of exponential random variables as in Section~\ref{model}. Doing these, we obtain
	\begin{align}
	E[Y_I] =&  \frac{M}{\lambda}H_n \label{thm1_eq1st} \\
	E[Y_{II}] =& \frac{n}{M^3\tilde{\lambda}}H_M \\
	E[Y_{III}] =&  \frac{M}{\lambda}H_{\frac{n}{M}}  \\
	E[Y_I^2] =&  \frac{M^2}{\lambda^2}H_n^2 + \frac{M}{\lambda^2}G_{n} \label{E[Y_I^2]} \\
	E[Y_{II}^2] =&  \frac{n^2}{M^6\tilde{\lambda}^2}H_M^2 + \frac{n}{M^5\tilde{\lambda}^2}G_{M} \\
	E[Y_{III}^2] =&  \frac{M^2}{\lambda^2}H_{\frac{n}{M}}^2 + \frac{M}{\lambda^2}G_{\frac{n}{M}}
	\end{align}
Lastly, we need to calculate $E[Z]$ where the random variable $Z$ is the additional amount of time after Phase~II ends until the destination node receives the update. Let us take an S-D pair $(s,d)$ where source node $s$ is from cell $j+1$. In Phase~III, $d$ has to wait for all other $j$ mega packets from the first $j$ cells to be distributed among the nodes. When its turn comes, $d$ just needs $X$ amount of time to get its packet. Then, $d$ has $Z = \sum_{i=1}^{j} \bar{\bar{U}}_i + X$. Taking expectation on $\bar{\bar{U}}$, $j$ and $X$ by noting their mutual independence we get
\begin{align}
E[Z] \!=\! \left(\frac{1}{M}\! \sum_{j=0}^{M-1}j \right)\!E[\bar{\bar{U}}] +E[X] = \frac{M-1}{2\lambda}H_{\frac{n}{M}} + \frac{1}{\lambda} \label{thm1_eqlast}
\end{align}
Using (\ref{thm1_eq1st})-(\ref{thm1_eqlast}) in (\ref{age_propScheme}) yields the expression.
\end{Proof}

Having derived the expression for the average age $\Delta$ of an S-D pair, we are now ready to work with large $n$.

\begin{theorem} \label{thm2}
For large $n$ and with $M = n^b$, where $0 < b \leq 1$, average age $\Delta$ in Theorem~\ref{thm1} approximately becomes,
\begin{align}
\Delta \approx& \frac{n^b}{\lambda} \log n + \frac{n}{n^{3b} \tilde{\lambda}} b\log n+ \frac{n^b-1}{2\lambda}(1-b)\log n+\frac{1}{\lambda} \nonumber \\ &+
\frac{\frac{n^{2b}}{\lambda^2}(\log n)^2 +\frac{n^b}{\lambda^2}\frac{\pi^2}{6}}{2\left(\frac{n^b}{\lambda}\log n+\frac{n}{n^{3b}\tilde{\lambda}}b\log n + \frac{n^b}{\lambda}(1-b)\log n\right)} \nonumber\\ &+ \frac{\frac{n^2}{n^{6b}\tilde{\lambda}^2}b^2(\log n)^2+ \frac{n}{n^{5b}\tilde{\lambda}^2}\frac{\pi^2}{6}}{2\left(\frac{n^b}{\lambda}\log n+\frac{n}{n^{3b}\tilde{\lambda}}b\log n + \frac{n^b}{\lambda}(1-b)\log n\right)} \nonumber \nonumber\\ &+
\frac{\frac{n^{2b}}{\lambda^2}(1-b)^2(\log n)^2 +\frac{n^b}{\lambda^2}\frac{\pi^2}{6}}{2\left(\frac{n^b}{\lambda}\log n+\frac{n}{n^{3b}\tilde{\lambda}}b\log n + \frac{n^b}{\lambda}(1-b)\log n\right)} \nonumber \\ &+
\frac{\frac{n}{n^{2b}\lambda\tilde{\lambda}}b(\log n)^2 + \frac{n^{2b}}{\lambda^2}(1-b)(\log n)^2}{\left(\frac{n^b}{\lambda}\log n+\frac{n}{n^{3b}\tilde{\lambda}}b\log n + \frac{n^b}{\lambda}(1-b)\log n\right)} \nonumber\\ &+
\frac{\frac{n}{n^{2b}\lambda\tilde{\lambda}}b(1-b)(\log n)^2}{\left(\frac{n^b}{\lambda}\log n+\frac{n}{n^{3b}\tilde{\lambda}}b\log n + \frac{n^b}{\lambda}(1-b)\log n\right)}
\label{thm2_res}
\end{align}
\end{theorem}

\begin{Proof}
The expression follows upon substituting $M = n^b$ in (\ref{thm1res}) and noting that for large $n$, we have $H_n \approx \log n$. Further, $G_{n}$ is monotonically increasing and converges to $\frac{\pi^2}{6}$. Since we have $M=n^b$, as $n$ grows large $M$ does too, resulting in $H_M \approx b\log n$ and $G_{M}$ converging to $\frac{\pi^2}{6}$.
\end{Proof}

\begin{theorem} \label{thm3}
For large $n$, and for $\frac{1}{4} \leq b \leq 1$, the average age of an S-D pair $\Delta$ given in (\ref{thm2_res}) reduces to,
\begin{align}
\Delta \approx c n^b\log n
\end{align}
with a constant $c$. That is, age is $O(n^b\log n)$, for $\frac{1}{4} \leq b \leq 1$.
\end{theorem}

\begin{Proof}
By analyzing the result of Theorem~\ref{thm2} we note that the first and third terms are $O(n^b\log n)$, and the second term is $O(n^{1-3b}\log n)$, and fourth term is a constant independent of $n$. Assuming $b \geq 1-3b$, the fifth term becomes
\begin{align}
\frac{n^{2b}(\log n)^2\left(\frac{1}{\lambda^2}+\frac{\pi^2}{6n^b\lambda^2(\log n)^2}\right)}{n^b\log n\left(\frac{2}{\lambda}+\frac{2b}{\tilde{\lambda}n^{4b-1}} + \frac{2(1-b)}{\lambda}\right)}
\end{align}
which is $O(n^b\log n)$. Continuing similarly for the remaining terms yields the result.
\end{Proof}

Thus, the proposed transmission scheme, which involves intra-cell cooperation and inter-cell MIMO-like transmissions, allows the successful communication of $n$ S-D pairs, and achieves an average age scaling of $O(n^{\frac{1}{4}}\log n)$ per-user. Noting that $\log n$ is negligible compared to the $n^{\frac{1}{4}}$ term, we state our result more succinctly as $O(n^{\frac{1}{4}})$.

\section{Note on Phases~I and~III} \label{note_phaseI}
We use the protocol model introduced in \cite{Gupta00} to model the interference such that two nodes can be active if they are sufficiently spatially separated from each other. In other words, we allow simultaneous transmissions provided there is no destructive interference caused by other active nodes. Suppose that node $i$ transmits its update to node $j$. Then, node $j$ can successfully receive this update if the following is satisfied for any other node $k$ that is simultaneously transmitting,
\begin{align}
d(j,k) \geq (1+\gamma)d(j,i) \label{prot_model}
\end{align}
where function $d(x,y)$ denotes the distance between nodes $x$ and $y$ and $\gamma$ is a positive constant determining the guard zone.

In the proposed scheme, we assume that the intra-cell transmissions that take place in Phases~I and~III work in parallel across the cells, whereas inter-cell transmissions that take place during MIMO-like transmissions in Phase~II work in sequence. In order to implement parallel intra-cell communications in Phases~I and~III, we follow a 9-TDMA scheme as in \cite{Ayfer07}. Specifically, $\frac{n}{9M}$ of the total $\frac{n}{M}$ cells work simultaneously so that Phases~I and~III are completed in 9 successive sub-phases. Using the protocol model, cells that are at least $(1+\gamma)r\sqrt{2}$ away from a cell can operate simultaneously during these phases, where $r=\sqrt{SM/n}$ is the length of each square cell and $S$ is the network area. Noting that there are at least two inactive cells in between two active cells under a 9-TDMA operation, this scheme satisfies (\ref{prot_model}) if the guard zone parameter $\gamma \leq \sqrt{2}-1$. Since 9 here is constant and valid for any $n$, it does not affect the scaling results.

\section{Conclusions} \label{conc}

We have studied the scalability of age of information in a large wireless network of fixed area with randomly paired $n$ source-destination pairs that want to update each other. We have proposed a three-phase transmission scheme which uses local cooperation between nodes and mega update packets to achieve an average age scaling of $O(n^{\frac{1}{4}})$.

Our scheme divides the network into $\frac{n}{M}$ cells of $M$ nodes each. The first and third phases include intra-cell transmissions and take place simultaneously across all cells. The second phase includes inter-cell transmissions and therefore during this phase cells operate one at a time. We create mega update packets in Phase I such that each mega packet includes all $M$ update packets to be transmitted from that cell. In the second phase, all $M$ nodes of a cell transmit this mega update packet to respective destination cells. Thus, by utilizing these mega update packets, we serve all $M$ source-destination pairs at once. Finally, in the third phase received mega update packet is relayed to its actual intended destination nodes. This node then extracts its update from the mega packet.

\bibliography{lib}
\bibliographystyle{unsrt}

\end{document}